**Silo Music and Silo Quake: Granular Flow Induced Vibration**

Benson K. Muite, Shandon F. Quinn and Sankaran Sundaresan*
School of Engineering and Applied Science,
Princeton University,
Princeton, NJ 08544, USA

And

K. Kesava Rao
Department of Chemical Engineering,
Indian Institute of Science,
Bangalore, India

Keywords: Silo quake, silo music, stick-slip flow, resonance, creep, granular discharge

*Corresponding author. Tel.: +1-609-258-4583; fax: +1-609-258-0211

*E-mail address*: sundar@princeton.edu (S. Sundaresan)

**Abstract**

Acceleration and sound measurements during granular discharge from silos are used to show that silo music is a sound resonance produced by silo quake. The latter is produced by stick-slip friction between the wall and the granular material in tall narrow silos. For the discharge rates studied, the occurrence and frequency of flow pulsations are determined primarily by the surface properties of the granular material and the silo wall. The measurements show that the pulsating motion of the granular material drives the oscillatory motion of the silo and the occurrence of silo quake does not require a resonant interaction between the silo and the granular material.

# 1    Introduction

The discharge of granular materials from silos is often characterized by vibrations or pulsations of the silo, termed 'silo quake', and a loud noise, termed 'silo music' [1-8]. Both of these are undesirable as silo quake may cause structural failure and silo music is a source of noise pollution. Unfortunately, the numerous conflicting studies published in the literature [1-8] do not give the silo designer a simple model to understand the physical processes that cause the pulsations, and to guide silo design or modification that would prevent the pulsations or at least minimize their effect. The purpose of this study is to investigate the cause of the noise and the pulsations, and the interaction between the motion of the granular material and the motion of the structure.

Several studies of the discharge of granular material from silos have noted fluctuations in discharge rate and the production of noise and vibration [1-8]. The top of the granular material has been observed to move in discrete steps even though the discharge from the bottom of the silo was continuous [4,6]. For smooth-walled, tall, narrow silos, pulsations occurred during both mass and mixed flow. The pulsations were observed to stop at a critical height of granular material in the silo [1,3]. Methods suggested for preventing pulsations include roughening the walls in the transition zone between the bunker and the orifice [1-3] and placement of inserts along the silo walls [4].

In an early study, Phillips [6] observed the motion of sand in a tube, which had a glass face, and was closed at the lower end by a flat bottom having a central orifice. When the orifice was opened, the sand in the upper part of the tube moved downward intermittently in jerks. Phillips noted, "when the flow begins, a curious rattling sound is heard which changes to a distinct musical note". He also did experiments in which the tube was first partly filled with mercury and then filled with sand. Once again, the free surface of the sand descended intermittently when the mercury was allowed to flow through the orifice. He observed that the length of the column of sand increased by about 2% during the 'stick' phase. Further, the motion of the granular material caused the wall of the tube to vibrate. Thus both silo music and silo quake occurred in his experiments, and he suggested that the stick-slip motion of the sand may be responsible for these phenomena.

Some recent studies have suggested that the pulsations are due to a resonant interaction between the granular material and the silo structure [1,7]. Unfortunately, these studies have been done in complicated systems where it is difficult to uncouple the effects of the vibrating structure from the pulsation of the granular material. In particular, understanding the source of measured tube wall vibration frequencies without a good numerical or theoretical model is difficult [9]. Further, a study has been conducted where the measured natural frequency of free vertical oscillation of the structure was significantly larger than the pulsation frequency, suggesting resonance did not occur [4]. Tejchman [1] also noted that the magnitude and presence of the flow pulsations was influenced by environmental factors such as temperature and electrostatic effects, which suggests that resonant interaction may not be required for pulsations to occur.

Wensrich [5] has proposed that these pulsations are due to compression and dilation waves in the granular material, which are created by stick-slip motion between the granular material and the silo walls. He found that at the very low discharge rates examined in his study (corresponding to average particle velocities in the cylindrical tube of 0.4 mm/s or smaller), the pulsation frequency was inversely proportional to the discharge rate. Pulsations have also been observed in funnel flow bunkers, where the granular material at the walls does not slip during discharge [4]. Wensrich [5] has suggested that the pulsation creation mechanism is entirely different in funnel flow, but does not give evidence to support his conjecture.

Hardow et al. [4] have suggested that the pulsations are due to the rapid acceleration and deceleration of the granular material in the bin section, caused by the stress fluctuations in the granular material in the hopper section. As the granular material in the hopper region deforms, there are periods where the mass of granular material in the bin is not supported and collapses in a downward step creating a large impulse, which shakes the silo structure. This study observed pulsations during core flow in a silo that was 6 m high, 0.6 m deep and 1.2 m wide, and hence the flow kinematics were considerably different from those in tall narrow silos.

Finally, Moriyama and Jimbo's [2] findings suggest that the magnitude of the pulsations is determined by how the granular material changes from a compressed state in the bunker to a dilated state in the hopper. Moriyama and Jimbo [2] also found that the likelihood a silo will discharge with pulsations is dependent on the method used to fill the silo. They did not propose a physical mechanism to explain their observations.

The aim of the present study, which is largely experimental, is to obtain a mechanistic understanding of silo music and flow pulsations. Through a combination of sound, bed height, and acceleration measurements, it is shown that silo music is driven by the stick-slip pulsating motion of the granular material during discharge and is associated with a sound resonance in the air column above the bed. Since previous studies have suggested a resonant interaction between the granular flow and silo structure as the possible cause for the pulsating motion of the granular material [1,7], a model silo system having a single dominant natural (vibration) frequency for vertical oscillations, which can be varied systematically was designed for this study. This has allowed a study of the interaction between the granular flow and the silo structure, which conclusively demonstrates that a resonant interaction between the granular material and the silo structure is not required to establish pulsating flow and silo music. Different wall and granular materials have been used to probe their role on flow pulsations and silo music during silo discharge.

## 2   Experimental Method

In this section, a description of (a) the model silo used in the study, which behaves like an oscillator with a single degree of freedom, (b) the properties of the granular and silo materials, (c) the characteristics of the accelerometers used to measure the motion of the silo and the granular material, and (d) the procedure used to obtain the sound measurements are provided.

Aluminum, steel, and acrylic tubes, open at the top and covered at the bottom with a flat acrylic plate having a concentric orifice (see below for details of this plate), were used as silos. A number of experiments were conducted using silos resting on supporting springs (see figure 1), which in turn were attached to a steel frame that was rigidly connected to the laboratory walls. The silo was also equipped with rollers and sliders, which were attached to the steel frame. These allowed vertical oscillation of the silos and restricted lateral motion. The supporting springs had spring constants ranging from 4 to 2265 N/mm. Experiments were also done using an aluminum block in place of the spring, or simply bolting the silo directly to the supporting steel frame – these configurations afforded the two largest natural frequencies for vertical silo oscillation reported in this study. Properties of the tube and granular materials are listed in tables 1 and 2, respectively. Photographs of the granular materials, obtained using a microscope, are shown in figures 2, 3 and 4. The granular materials did not exhibit squeaking or booming when sheared. The temperature and humidity were recorded in each experiment. The temperature varied between 20°C and 25°C (from one day to another), and the relative humidity between 18 and 40%. During experiments with each tube and granular material combination (which lasted a few hours) the humidity variation was within 5% and the temperature variation was within 2°C.

Since structural accelerations were measured, care was taken to ensure that the apparatus had a single controlled dominant natural frequency for vertical oscillations. The natural frequency of vertical oscillations was changed by using different springs between the silo and the steel frame. In these experiments, the metal tubes were chosen to have large natural frequencies, so that when the structure was excited by a broadband impulse in the vertical direction, the dominant natural frequency was due to the mass-spring system, and not the intrinsic vibration of the tube. For further details, see Muite [10]. The acrylic tube had a natural frequency of 350 Hz, which while not as large as the metal tubes was still several times greater than the pulsation and lowest structural natural frequencies.

Acrylic plates with centrally located orifices (with diameters between 13 and 25 mm) were bolted to a 1.21 Kg aluminum flange, which was screwed on the bottom of the tubes. To fill the silo, the orifice at the bottom of the tube was first sealed with a piece of duct tape. The granular material was poured into the silo through a funnel placed at the top of the tube. Stripping away the duct tape seal over the orifice initiated discharge. The mean discharge rate was measured using a stopwatch. For the acrylic tube, the height of material in the silo could also be measured during discharge to confirm that the discharge rate was constant with time. To ensure that the tubes had reached a steady state of wear, the granular material was discharged several times through the same tube before final measurements were taken. Steady state wear was reached when repeatable granular material acceleration measurements could be taken.

Accelerations were measured both in the granular material and on the silo structure. Vertical accelerations inside the granular material were measured using a unidirectional Kistler 8774A50 low-impedance ceramic shear accelerometer with an output sensitivity that deviated less than 1.5 % for frequencies between 10 Hz and 10 KHz. The accelerometer was embedded approximately 50 mm below the top surface of the granular material. This depth ensured that during discharge the accelerometer was held upright by the granular material and was still shallow enough that the acceleration could be measured for the bulk of the discharge. As the granular material discharged, the accelerometer cable was carefully fed into the silo to ensure that the cable did not affect the motion of the accelerometer. This accelerometer had a range of $\pm$ 500 m/s$^2$ and was accurate to within $\pm$ 5 m/s$^2$. It had a diameter of 8 mm, a length of 26 mm, and a mass of 4 g, and so was considerably larger than a sand grain. However, the wide frequency response allowed better time resolution of the bulk granular material acceleration than would be possible with smaller accelerometers of comparable cost.

Silo structural vibrations were measured using a Kistler 8784A5 low-impedance ceramic shear accelerometer which has a greater sensitivity but a smaller range than the accelerometer used to measure granular material accelerations. To measure vertical accelerations, this accelerometer was wax mounted on the flange at the bottom of the silo. This accelerometer had a sensitivity that varied by less than 0.5 % for frequencies between 10 Hz and 6 KHz. It had a range up to $\pm$ 50 m/s$^2$, an accuracy of $\pm$ 0.5 m/s$^2$, and a mass of 21 g.

The accelerometer output was sent through a Kistler 5118B2 signal conditioner to a Measurement Computing PCI-DAS1002 data acquisition card on a 400 MHz Pentium II computer. The sampling rate on the data acquisition card was 20 KHz. For both accelerometers, the manufacturer supplied calibration was used to convert the accelerometer voltage output to acceleration. The accelerometers could not be used simultaneously since only one data acquisition system was available.

The bulk of the sound measurements were made in an apparatus made from an acrylic tube for which the resonant frequency of the structure was not well controlled [11]. However, several measurements were then repeated in the experimental setup used for the acceleration measurements to check that the same results were obtained. In these experiments an omnidirectional Optimus 33-3026 lapel microphone with a constant amplitude response for a frequency range between 30 Hz and 15 KHz was used to collect the sound data through a sound card on a personal computer. During discharge, the sound was recorded and a discrete Fourier transform of one second of sound data was used to determine the dominant frequency as a function of time during discharge. In the acrylic tube, the time at which the top of the granular material crossed a marked height in the tube during discharge was also recorded using a stopwatch. From these measurements the height of the granular material as a function of time since discharge started was found.

## 3 Results

To present the trends that are useful in determining the pulsation mechanism, sound measurements are only presented for sand discharging from the acrylic tube and acceleration measurements are shown primarily for crushed glass particles discharging from the aluminum and acrylic tubes. Measurements for glass beads are also presented. For more measurements of silo music produced by sand discharging from an acrylic tube see Quinn [11] and for more acceleration measurements during granular flow pulsations see Muite [10]. The variation of the pulsation frequency with the natural frequency of vertical oscillations of the silo is examined for all tube and granular material combinations, except for the plain steel tube, as pulsations did not occur in this tube. Silo pulsations also did not occur when sand was discharged from the aluminum tube, but did occur when sand was discharged from the acrylic and galvanized steel tubes.

### 3.1 Sound measurements

Figure 5 shows the sound intensity level as a function of time for discharge of sand from the acrylic tube. The discharge lasted for 51 s and as shown in the figure, silo music occurred for approximately half of this time. Figure 6 shows a typical power spectrum for the sound measurements during discharge (determined by analyzing data obtained over a 1 s time interval). There are three types of prominent peaks. The first peak is at a frequency of approximately 40 Hz and it will be shown later that this is the pulsation frequency of the granular material and silo combination. The second peak corresponds to the resonant frequency for the air column above the tube (this resonance is well documented and a good account can be found in Rayleigh [12]). At the time the data shown in the figure was collected, this frequency was 200 Hz. The fact that this peak represents a resonance frequency is demonstrated in Figure 7, which shows the quarter wavelength corresponding to this frequency as a function of time since the beginning of discharge. The wavelength, $\lambda$, is found from the relationship $\lambda = c/f_a$ where $c$ is the speed of sound in air and $f_a$ is the frequency. Also shown is the height of the air column above the sand in the tube. This figure shows that the dominant quarter wavelength and the height of the air column are the same confirming the resonant behavior. It is clear from the *quarter* wavelength, that this resonance corresponds to a standing wave mode with a node at the granular material surface, and an anti-node at the open end of the tube (as the open end of the tube cannot be a node). Figure 6 also shows a number of other peaks at higher frequencies, which are simply the odd harmonics of the fundamental (lowest) resonance frequency of the air column.

### 3.2 Determination of the natural frequency for vertical oscillations of the silo

Figure 8 shows a typical measurement to determine the natural frequency of vertical silo oscillations. In these measurements, the silo was filled with granular material and the orifice closed. The base of the filled silo was then struck with a soft mallet and the resulting acceleration during free oscillations recorded. The dominant natural frequency of the silo structure was found either by using the largest peak in the power spectrum of the acceleration, or by counting the number of free oscillations during a specified time directly from the acceleration measurements. The two measurements gave essentially the same results; however, when the natural frequency was less than 30 Hz, counting the number of oscillations in a specified time gave a more accurate measurement of the natural frequency than locating the center of the broad peak obtained from the power spectrum. Similarly when the natural frequency was large, the power spectrum was a better indicator of the natural frequency because an unambiguous sharp peak could be located, while the acceleration vs. time trace showed rapidly decaying oscillations which were not easy to count. For spring constants, $k < 1000$ N/mm, the observed natural frequency was approximately equal to $f_n = (1/2\pi)(k/m)^{1/2}$, where $m$ is the oscillating mass and $f_n$ is the theoretical natural frequency for a spring mass system. For $k > 1000$ N/mm, the natural frequency of vertical silo motions was significantly less than $f_n$, possibly because of flange deformations, which reduced the effective stiffness of the system. This effect was important for natural frequencies greater than 25 Hz.

### 3.3 Acceleration measurements during discharge

Figure 9 shows measurements of the vertical acceleration of the silo when crushed glass was discharged through a 1.9 cm orifice. The accelerometer was mounted on the base of the silo. Once flow started, there was a period of pulsations during which the silo experienced large negative accelerations towards the earth. Half way during the pulsations, the magnitude of the negative pulsations suddenly doubled. After the granular material fell below a critical level, the pulsations stopped, and the silo structure experienced only small accelerations until the flow ended. While the pulsations occurred regularly, this doubling of the pulsation magnitude was not always repeatable. It is not clear what changes in the flow resulted in these changes in the magnitude of the pulsations since the basic setup was unchanged from run to run.

The close up of the acceleration measured during pulsations shown in figure 10 reveals that the periods of large negative accelerations were short compared to the gradual rebound after each pulsation. Over this time scale, the pulsations had a very reproducible and steady frequency, but the absolute magnitude of the maximum acceleration varied from pulse to pulse.

Figure 11a shows measurements obtained with the accelerometer buried in the granular material. The flow conditions were the same as in figure 9, except that the orifice diameter was 1.3 cm instead of 1.9 cm. Also as in figure 9, negative accelerations are to the earth. Figure 11a shows that large positive accelerations occurred in the granular material during pulsations, while figure 9 shows that the silo experienced large negative accelerations. The two figures show that during each pulsation, the granular material fell a short distance and impacted the tube wall and flange bottom.

The close-up of the pulsations in figure 12a shows that despite the difference in flow rate, the pulsation frequency of 30 Hz is the same for discharge of granular material through a 1.3 cm orifice (figure 10) and through a 1.9 cm orifice (figure 12a). Figures 10 and 12a also show that the pulsation frequency in the granular material is the same frequency with which the silo moves, suggesting that the motion of the granular material drives the motion of the silo. Figure 12a also shows that each pulsation was followed by a negative acceleration within the material and then a second large positive acceleration, after which the acceleration of the granular material was close to zero until the next pulsation. Figure 13a shows a power spectrum for the acceleration measured during 1 second of pulsations in the crushed glass. It has a peak at the pulsation frequency of 30 Hz followed by a flat band region between 200 and 1000 Hz after which the power spectrum decays.

Figures 11b, 12b, and 13b are similar to figures 11a, 12a and 13a, but are for glass beads discharging through a 1.9 cm orifice from a silo resting on a 22 N/mm spring. The pulsations again stop at a critical height and the individual pulsations can be seen in figure 11b. The nature of each pulsation for glass beads (figure 12b) is a little different than for the crushed glass (figure 12a) and this is reflected in their power spectra, (figures 13a and 13b). Both spectra have the same high frequency decay for frequencies above 1000 Hz; however, for frequencies below 1000 Hz the glass beads have a larger number of distinct harmonics than the crushed glass. The crushed glass power spectrum is typical of white noise with a high frequency cutoff, while the glass bead power spectrum is typical of a signal produced by a well correlated periodic, but non-sinusoidal function [13].

In figures 10, 12a, and 12b the maximum downward accelerations of the silo and particles are roughly comparable, whereas the maximum upward acceleration of the granular material is significantly greater than that of the silo. This suggests that during each pulsation, the granular material slips past the silo walls and is forced to rest over a very short time period. This impact creates a shock wave which travels through the granular material and is recorded as the large upward acceleration. The granular material and silo then move together so that the resulting accelerations are of similar magnitude.

### 3.4  Dependence of the pulsation frequency of the granular material on the natural frequency of vertical silo vibrations

Figures 14, 15, and 16 show the variation of the pulsation frequency of the granular material for different granular material and silo wall combinations, as a function of the natural frequency of vertical oscillations of the silo. The pulsation frequency was determined in two ways. First, by counting peaks above a threshold value when the accelerometer was placed in the granular material and second by finding the largest component in the power spectrum when acceleration was measured on the base of the silo. These two methods were used because the power spectrum for acceleration in the granular material gave many harmonics and the lowest frequency did not always have the largest power. Similarly counting silo oscillations was not as precise as locating the single sharp peak in the power spectrum. However, in most cases, the results from the two methods were comparable. The pulsation frequency was determined for each second of flow pulsations and an average pulsation frequency during pulsating discharge obtained. The standard deviation in the average frequency measured during a single discharge was typically less than 10%.

When the silo had a free oscillation frequency below 25 Hz, the pulsation frequency had no dependence on the natural frequency of the silo as shown in figures 14, 15, and 16. Figure 15 also shows that doubling the orifice diameter and hence increasing the discharge rate by a factor of 6 had a negligible effect on the pulsation frequency (doubling the orifice diameter gives a six-fold increase in discharge rate in a silo of constant cross sectional area, because the discharge rate is proportional to the orifice diameter to the power 2.5, see Nedderman [14] for more details). The pulsation frequency was solely determined by the silo wall and granular material properties. For silo free oscillation frequencies above 25 Hz, the pulsation frequency had a positive correlation with the free oscillation frequency for all granular material tube wall combinations that pulsated, except for the acrylic and crushed glass combination. These figures show that the pulsations were not due to a resonant interaction with the silo structure, even though the motion of the silo can couple with the motion of the granular material and influence the frequency of pulsations. The three figures show that glass beads have similar frequency behavior in the acrylic and aluminum silos, but they have a higher pulsation frequency in the galvanized steel silo. Crushed glass has a lower pulsation frequency in the aluminum silo as compared to the acrylic silo, and a higher pulsation frequency in the galvanized steel silo compared to the acrylic silo. Since sand did not pulsate during discharge from the aluminum silo, no data points are shown.

### 3.5  Critical Height

The critical height is taken as the height of the granular material above the base of the silo at which pulsations stop. The time at which this occurred was recorded from the acceleration measurements, and as the discharge rate was independent of time, the critical height could be calculated. This method gave critical heights that were in agreement with direct measurements made for the transparent acrylic silo. The critical height did not vary by more than 0.1 m when the orifice diameter and spring constant were changed as long as the resonance frequency of vertical oscillations of the filled tube was less than 25 Hz. For frequencies above this, the variation of critical height with silo and granular material properties was not closely examined.

For all the granular materials, the critical heights for the acrylic silo were smaller than those for the aluminum and galvanized steel silos (table 3). In the aluminum silo, crushed glass had a significantly smaller critical height than glass beads. In both the galvanized steel and acrylic silos, all three granular materials had similar critical heights.

In the aluminum silo, an experiment was performed where the top of the granular material was loaded with a known weight after the silo had been filled. The weights were placed on top of the center of the granular material in the filled silo and did not touch the walls of the silo. The critical height and pulsation frequency were measured using time and acceleration measurements during discharge. The pulsation frequency was independent of the overload. As shown in figure 17, the critical height for glass beads was insensitive to the imposed overload, whereas that for crushed glass decreased linearly with the imposed overload. Experiments in the acrylic silo gave similar results.

## 4   Discussion

The goals of this study were to investigate the basic mechanism that causes silo music and silo quake, and to understand the interaction between the silo structure and the granular material during silo quake. In this section, a summary of the experimental results is given and a mechanism for the production of the pulsations is suggested. The results are then compared with those obtained in previous work on pulsating granular materials and some suggestions for further work are made.

The main observations of this investigation are:

1) In tall, narrow silos, silo quake is due to the frictional interaction between the granular material and the silo walls.
2) Silo music is created by silo quake and is a quarter wave sound resonance in the air column above the granular material.
3) Silo quake can occur without silo structural resonance.
4) A stick-slip regime has been observed where the motion of the granular material drives the motion of the silo. In this regime, the pulsation frequency has a narrow band distribution and does not depend on the discharge rate.
5) The upward acceleration of the granular material can reach up to 500 m/s$^2$, while the maximum acceleration of the silo was observed to be less than 15 m/s$^2$.

Before discussing pulsating flow, it is worth reviewing the generally understood kinematics of the discharge of granular material from a bin or hopper. Experiments show that in a tall, flat-bottomed cylindrical bin, with walls having a lower friction coefficient than the internal friction angle of the granular material, there is a region of plug flow at the top of the full silo. As the silo empties, the size of the plug flow region decreases and eventually all of the flowing material is in converging flow. The discharge rate from the bin is independent of the height of material in the bin, provided the height is greater than a few multiples of the diameter of the silo [20] and scales as $g^{1/2}D^{5/2}$, where $D$ is the orifice diameter.

Radiographic studies of slow dense granular flow in model bunkers show that velocity discontinuities exist at the transition from the bin to the hopper [15]. Measurements in a discharging bunker indicate that there exists a dynamic arch at the transition where the nature of the material flow changes from one without deformation (above the arch) to one where material deforms (below the arch) as it approaches the orifice [16]. Pressure measurements [2,4,7] near the transition from the bin to the hopper indicate that there is also a stress discontinuity [16] and that there are large pulsating stresses, which correspond to the cyclical formation and breakage of the dynamic arch. This pulsating behavior only occurs for dense assemblies [16,17].

These experiments were done in a flat bottomed silo, whereas the experiments that identified the dynamic arch [16] were conducted in a bunker where the bin to hopper transition determined the location of the dynamic arch. In the silo used in this study, the stagnant material adjacent to the orifice creates a hopper-like region. So the discharge from a flat bottomed silo may be expected to show many of the features observed in bunkers. If the density of the material in the plug flow region above the dynamic arch is high, it must dilate as it crosses the arch in order to deform in the hopper-like region.

Stick-slip motion in granular materials has also been studied in geometries other than silos. Nasuno et al. [18] performed simple shear tests with 70 - 110 µm glass beads and 100 - 600 µm sand. They observed stick-slip motion at low slip rates, which became continuous at very large slip rates. They also observed that at very small driving velocities, the period of fluctuations was inversely proportional to the driving velocity. For glass beads, the system fluctuated with near constant period, while for sand, the period varied stochastically. As the sliding velocity was increased, the period became independent of velocity, and finally at large sliding velocities, the motion became continuous. In the simple shear experiments done by Nasuno et al. [18], the spring constant connecting the driving piston to the sliding mass was varied. The spring constant influenced both the pulsation frequency and the critical driving velocity at which the pulsation frequency became independent of the driving velocity.

Nasuno et al. [18] also observed that a lengthy period of slow vertical dilation preceded rapid slip events in the horizontal direction. This dilation was carefully measured by Geminard et al. [19]. Their experiments suggest that in the shear zone particles climb slowly over each other. Once the particle is approximately 5% of a particle diameter over the particle below it, slip occurs and the top layer jumps forward before slowing down again and settling into another zone of particles [19]. In the experiments of Geminard et al. [19], the particle volume fraction was indistinguishable from the random close packed volume fraction of 63%.

These studies show that granular materials can undergo stick-slip motion, and that this can couple with the mechanical system (for example, a mass-spring system) in a complicated fashion, which depends on the system parameters. Together these studies suggest that stick-slip motion can occur in tall flat-bottomed silos during the discharge of granular materials. This leads to cyclical dynamic arch formation and breakage, which creates impulses that drive the silo structure. The stick-slip motion of the granular material can couple with the motion of the silo, but as shown by the results, the frequency of the stick-slip motion of the granular material need not be the same as the dominant natural frequency of the silo.

Armed with this background, one can compare the experimental observations in this study with those in previous studies to qualitatively explain the frequency dependence and the physical mechanism of stick-slip motion. As mentioned above, if the density of the material in the plug flow region above the arch is high, it must dilate as it crosses the arch in order to deform after it has passed the dynamic arch. This dilation process and stick-slip friction in the granular material or at the silo wall can create granular material pulsations which originate at the arch. Excellent reviews on stick-slip friction can be found in Bowden and Tabor [21], Krim [22] and Berman et al. [23], who discuss the various postulated stick-slip friction mechanisms. The mechanism of most relevance to the present study is adhesive stick-slip friction which occurs when there are slowly weakening, time-dependent forces between the sliding surfaces. The forces are due to interactions at the contacts between the sliding surfaces which weaken after a typical lifetime.

The dynamic arch is a force chain – that is, a fragile network through which stresses are transmitted [24]. This arch supports a large fraction of the pressure above it and transmits this force to the walls of the silo. The arch is fragile, and consequently when the material below it has discharged enough so that the arch is unsupported from below, the slow creep typically observed in adhesive stick-slip begins. After the critical creep time during which the surface forces weaken, complete slip occurs collapsing the arch and creating a new arch. The process then repeats, giving rise to periodic impulses as the granular material above the arch moves in steps. During pulsations, material above the arch does not move smoothly, however material at the orifice exits continuously. This is the pulsation process observed in the present study, where the discharge rate is fast enough that it does not affect the pulsation frequency unlike in Wensrichs' study [5].

In Wensrichs' experiments [5], the entire bottom of a cylindrical model silo was slowly lowered. There is no region of converging flow, but there is a region near the bottom where the granular material dilates as the piston descends. Here the arch may be regarded as the boundary between the dense and dilated material.

For the *slow* discharge rates examined by Wensrich [5] where the average velocity of descent through the tube was 0.4 mm/s or smaller, creep was not the pulsation frequency determining step, and the arch collapsed whenever the particles below the arch had dropped enough to lose contact with the arch. Consequently, it is entirely reasonable that the pulsation frequency was inversely proportional to the discharge velocity. Wensrich estimated the distance that the particles below the arch moved between pulsations was 3% of a particle diameter, which is comparable to the 5% of a particle diameter dilation distance required for slip to occur in stick-slip simple shear experiments with granular materials [19]; however, further studies are required to determine if the proportionality constant (i.e. critical distance) in this regime is always a small fraction of the mean particle diameter.

Our experiments clearly revealed that at *large* average velocities of descent through the tube (10 – 60 mm/s), the pulsation frequency is independent of both discharge velocity and weak forcing due to silo structure oscillations. Hence it is reasonable to associate the observed pulsation frequency with the characteristic creep time. This finding also differs from experimental results obtained from studies of stick-slip motion in spring slider mechanisms where the period depends on the mechanical properties of the spring system (such as the spring constant and sliding mass) and is not solely controlled by the interaction between the sliding surfaces [18,21].

The lack of dependence of the pulsation frequency on discharge rate observed in this study is quite peculiar, as it implies that the distance the granular material falls during each pulsation increases as the discharge rate increases.

Wensrich [5] observed that the acceleration produced by each quaking impulse grew with distance traveled by the wave carrying the information of the impulse from the dynamic arch to the top of the silo. In contrast, the granular material accelerations recorded in these experiments with the accelerometer a fixed depth below the free surface did not change appreciably as the bed height decreased during discharge.

The granular material accelerations measured by Wensrich [5] were less than 15 m/s$^2$, while those measured in this study were typically more than 100 m/s$^2$. Tejchman [1] also observed silo wall acceleration levels greater than 100 m/s$^2$. Nonlinear effects that come into play at large amplitude disturbances or simply the fact that flow occurred in this system in a different frictional regime may be responsible for the height independence of the acceleration at the large accelerations seen in this study.

It is interesting to contrast the acceleration power spectra obtained with crushed glass and glass beads to see the effect of particle shape on granular dynamics. The power spectra for acceleration measurements in the glass beads (figure 13b) showed many harmonics of the pulsation frequency before the high frequency decay region was approached. The power spectra for crushed glass (figure 13a) showed only a few harmonics, followed by a band limited white noise region and then a high frequency decay. This suggests that the glass beads showed a highly correlated distributed response to slip which originated at the arch. Crushed glass had a significantly less correlated response, quite possibly because of the heterogeneity in particle shape and particle contacts between them. This is consistent with the suggestions by Mair et al. [25] that smooth round particles have force chains that are stable over a narrow range of orientations whereas, rough particles such as the crushed glass have force chains that are stable over a wider range of orientations. Consequently, the force chains in the glass beads break in a highly correlated manner during a pulsation, whereas those in the crushed glass break in a less correlated manner.

Surprisingly, all the power spectra for acceleration measurements inside the granular material for all tube and granular material combinations that pulsated decayed for frequencies above 1000 Hz. The high frequency cutoff of 1000 Hz was neither due to any limitation of the accelerometer (which could measure frequencies up to 10 KHz) or the lowest natural frequency of the tube (which was varied in these experiments and did not affect the high frequency cutoff). It may be related to the tube diameter, or to the particle size and particle density, which were not varied in the experiments.

The hypothesized difference in stress chain behavior between smooth and rough particles suggested by Mair et al. [25] may also explain the difference in the value of the critical height when an overload is imposed on the granular material. For rough particles the critical height decreased linearly with imposed overload, whereas for smooth particles, the critical height was independent of the imposed overload. Since the force chains in a granular material composed of smooth spheres will have narrow directionality, the effects of the imposed overload will be transmitted to the side walls of the silo rapidly, and will not affect stress levels between the silo wall and the granular material a significant distance away from the overload. These force chains form a bridge so that the bulk of the overload is transmitted to the silo walls. For rough particles, the force chains are less concentrated because friction and asperity interlocking allows rough particles to transmit forces in a variety of directions without failure. Because bridging in the granular material is less effective, the imposed overload is not screened and its effects on the stress field can be transmitted further in to the granular material. Consequently, the critical height decreases because stresses at the arch are large and allow pulsations to occur for a longer time during discharge, in agreement with studies that slip-stick friction is dependent on the local stress level [21]. Further work examining the shear and wall normal stresses in silos for different shapes and distributions of particle sizes with varying overloads would help in verifying these suggestions and in obtaining appropriate constitutive relations to describe granular material stress fields macroscopically, an area which is the subject of current debate [26].

Mair et al. [25] found that particle shape influences granular material sliding characteristics and influences stress transmission through the granular material. The present study confirms these findings, since granular materials made of the same glass with similar sizes but different shapes had different pulsation frequencies in the aluminum and galvanized steel tubes. This is because frictional characteristics depend on the roughness, hardness, shape, and chemical composition of the surface. In particular, since glass and aluminum have comparable hardness, particle shape will be the dominant factor determining the difference in sliding characteristics of crushed glass and glass beads. Since crushed glass has sharp edges, it can dig in to the aluminum surface more than the smooth glass beads, and so the adhesive stick time for crushed glass in the aluminum silo is longer than for glass beads in the same silo. Acrylic surfaces are prone to stick-slip motion [27]. As acrylic is softer than all the particles used in this study, the acrylic surface can be expected to be the dominant factor in determining the adhesive relaxation time for the stick-slip motion, and indeed it was found that the pulsation frequencies for all granular materials are similar in the acrylic silo. Glass beads had a lower slip frequency in the galvanized steel silo than crushed glass due to the different pressure interactions. Steel is harder than glass and since the steel surface is smooth, abrasive mechanical wear is negligible and surface chemical interactions are more significant. Since the glass beads used are smooth on a 10 μm length scales, they have a larger surface area in contact with the steel tube compared to the crushed glass. The glass beads therefore have greater adhesive forces, thus a larger adhesive relaxation time, and hence pulsate less often than the crushed glass particles in the steel tube. This is in agreement with observations that surfaces that are smooth on small length scales are more likely to undergo adhesive stick-slip friction since sticking and sliding can occur on the entire surface. Conversely, surfaces that are rough at the micron length scale are less likely to have stick-slip friction since the asperities can lock and prevent slip occurring at all contact points. These results are in accord with findings by Tejchman [1], Moriyama and Jimbo [2], and Jahagirdar [3] that to prevent stick-slip friction at the silo walls and hence silo quake, mass flow silos should have rough walls.

Tejchman and Niedostatkiewicz [17] and Kmita [7] had suggested that silo quake is due to a resonant interaction between the dynamics of the silo structure and the motion of the granular material. Hardow et al. [4] showed that a resonant interaction is not required for quaking to occur. This study confirms that the pulsation frequency need not be the same as the dominant natural frequency of the silo structure. Hardow et al. [4], whose study did not examine a variety of granular materials or change the frictional characteristics of the silo walls, suggested that wall friction was not the cause of silo quake, a finding which this study has shown is not always correct. Since their study was for a core flow silo, they did not consider the possibility that stick-slip friction can occur at sliding surfaces inside the granular material, as shown in the study by Nasuno et al. [18,] and not just between the granular material and the silo wall. The suggestion by Hardow et al. [4] that the silo is driven by the motion of the granular material is confirmed by the results of this study.

## 5   Summary

This study has shown that stick-slip motion generates silo music and silo quake. Silo music is driven by the stick-slip pulsating motion of the granular material during discharge and is associated with a resonance in the air column above the bed. When the pulsating motion disappears, so does the silo music. For small values of the spring constant, the pulsation frequency differs significantly from the dominant natural frequency of vertical oscillations of the silo. Hence, silo quake is not due to a resonant interaction between the granular material and the silo structure.

Over the range of discharge rates studied here (equivalent to average velocities of descent through the tube of 10 – 60 mm/s), the pulsation frequency was independent of discharge velocity and *weak* forcing due to silo structure oscillations. The pulsation frequency is controlled by the slow creep during the *stick* phase of the pulsating flow, with slip occurring on a much faster time scale. This slow creep is determined by the frictional interactions between the particles and the tube wall.

When the natural frequency of vertical oscillations of the silo is large, resonance can set in and the flow pulsates at the same frequency as the column. In this regime, the duration of the creep may be cut short by the oscillations of the silo.

Both silo music and flow pulsations stopped abruptly when the bed height fell below a critical value. It was found that the critical height could be changed by placing an overload in the case of crushed glass, but not smooth glass beads. This may be rationalized, although only speculatively at this point, on the basis of the differences in the behavior of stress chains in columns.

## 6  Acknowledgements

We are grateful to Professor A. Smits for many helpful discussions. The authors also wish to thank C. Wensrich for providing a copy of his thesis in advance of publication. BKM was supported by a Princeton University Francis Upton graduate fellowship. SFQ acknowledges financial support provided by the Derek Lidow senior thesis fund.

**Table 1**
**Tube properties**

| Tube Material | Length (m) | ID (mm) | Wall Thickness (mm) | Surface Finish | Experimental Natural Frequency (Hz) |
|---|---|---|---|---|---|
| 6061-T6 Aluminum Alloy | 1.8 | 63.7±0.1 | 5.1±0.1 | Smooth | 3000 ±100 (Small peak at 1500) |
| Untreated Steel | 1.8 | 63.8±0.1 | 5.7±0.1 | Rough | 2900 ±100 (Small peak at 1400) |
| Galvanized Steel | 1.8 | 63.2±0.1 | 5.2±0.1 | Smooth | 2800 ±100 (Small peak at 1400) |
| Cast Acrylic | 1.5 | 63.5±0.1 | 6.4±0.1 | Smooth | 350 ±100 |

**Table 2: Granular Material Properties**

| Material | Supplier | Particle Size (μm) | Particle Density (g/cm$^3$) | Angle of Repose (°) |
|---|---|---|---|---|
| Crushed Glass | Potters Industries | 450±50 | 2.5±0.1 | 28±1 |
| Ballotini Impact Beads | Potters Industries | 480 ±60 | 2.5±0.1 | 26±1 |
| Washed and Ignited Standard Ottawa Sand | EMD Science | 400±100 | 2.7±0.1 | 33±1 |

**Table 3**

**Variation of critical height with silo wall and granular material properties**

| Silo wall material | Granular material | Critical height[a] (m) |
|---|---|---|
| Aluminum | Crushed glass | 0.9 |
| Aluminum | Glass beads | 1.3 |
| Acrylic | Crushed glass | 0.8 |
| Acrylic | Glass beads | 0.6 |
| Acrylic | Sand | 0.7 |
| Galvanized Steel | Crushed glass | 1.3 |
| Galvanized Steel | Glass beads | 1.4 |
| Galvanized Steel | Sand | 1.4 |

Note: [a] The accuracy of the critical height data is ± 0.1 m.

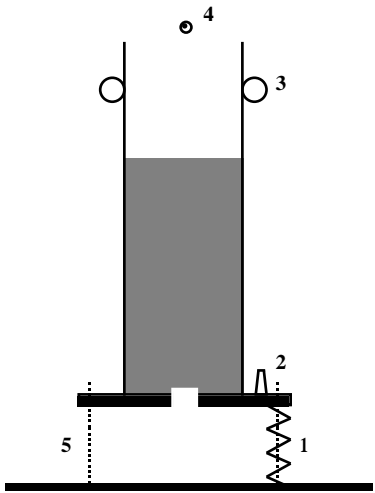

**Fig. 1:** Experimental setup for vertical acceleration and sound measurements. The numbers indicate (1) spring on positioning slider, (2) accelerometer, (3) positioning roller, (4) microphone, (5) positioning slider.

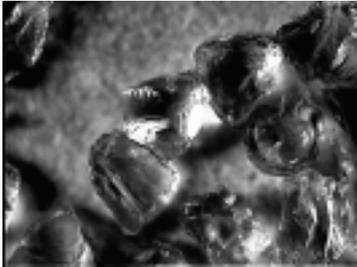

**Fig. 2:** A photograph of the crushed glass.

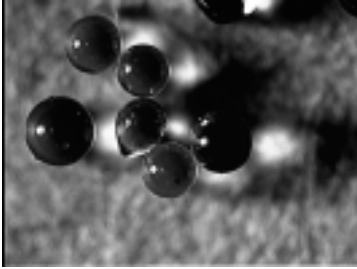

**Fig. 3:** A photograph of the glass beads.

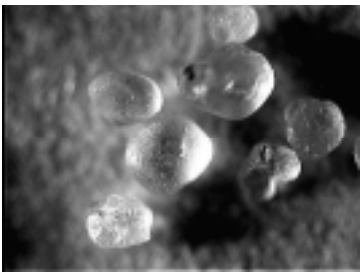

**Fig. 4:** A photograph of the sand particles.

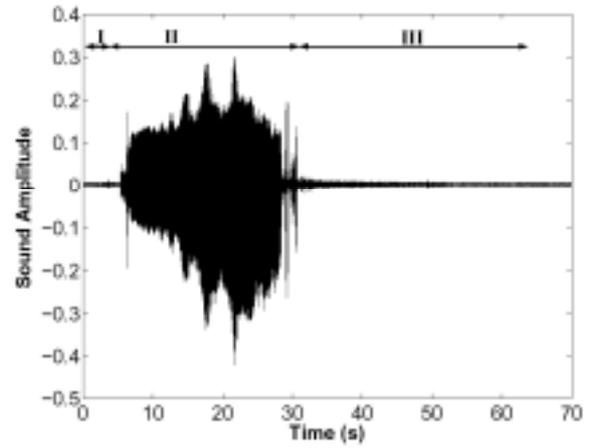

**Fig. 5:** Variation of sound level with time during discharge of sand from an acrylic tube of 76 mm outer diameter, wall thickness 3 mm, and having an orifice of diameter 19 mm. Indicated on the figure are; region I – no flow, region II – flow with pulsations and region III – flow after pulsations have ended.

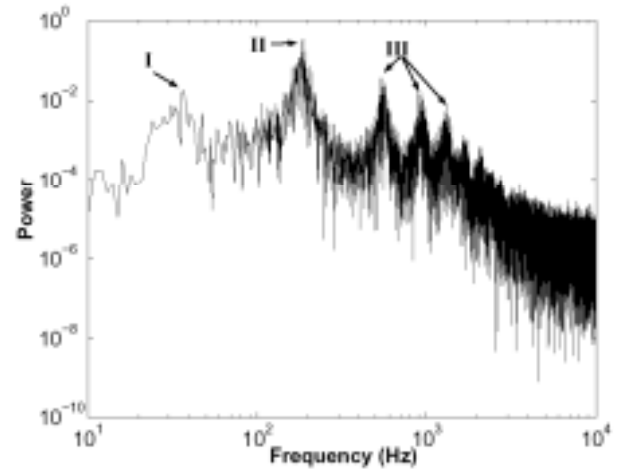

**Fig. 6:** Typical power spectrum for sound measurements during silo music when sand is discharged from an acrylic tube of 76 mm outer diameter, wall thickness 3 mm, and having an orifice of diameter 19 mm. Indicated on the figure are the pulsation frequency (I), the dominant sound frequency (II) and the higher harmonics of the dominant sound frequency (III).

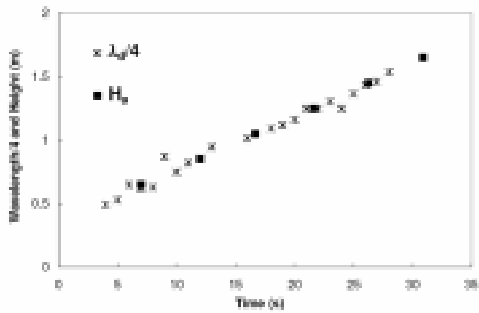

**Fig. 7:** Variation of the dominant quarter wavelength ($\lambda_d/4$) and the height of the air column ($H_a$) with time during discharge of sand from an acrylic tube of 76 mm outer diameter, wall thickness 3 mm, and having an orifice of diameter 19 mm.

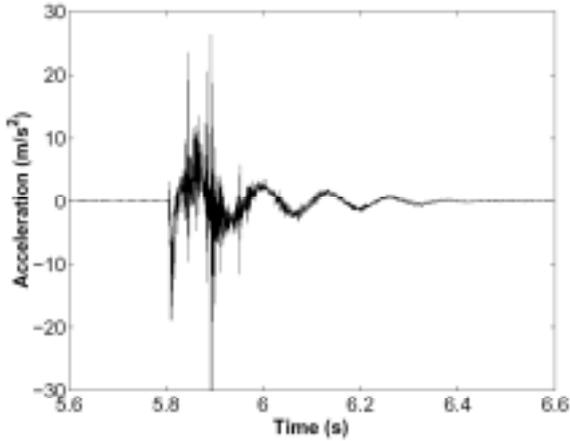

**Fig. 8:** Typical acceleration measurement for free oscillations of the filled aluminum silo on a 31 N/mm spring after it has been struck.

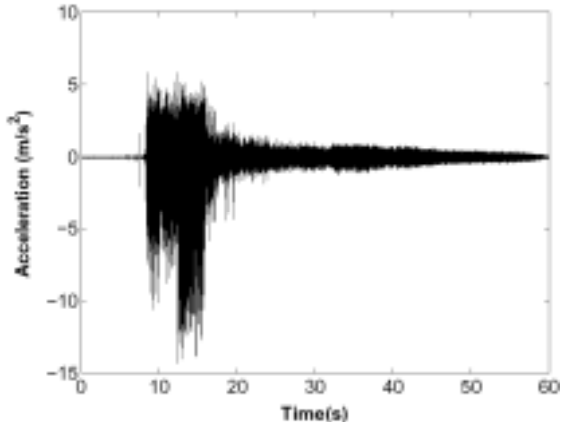

**Fig. 9:** Vertical acceleration measurements on the base of the aluminum silo during discharge of crushed glass through a 1.9 cm orifice. The silo was mounted on a 31 N/mm spring. Indicated on the figure are; region I – no flow, region II – flow with pulsations and region III – flow after pulsations have ended.

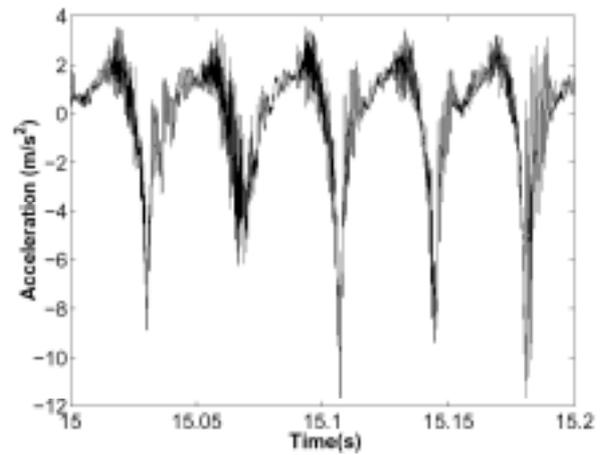

**Fig. 10:** Close-up showing individual pulsations measured by the accelerometer on the silo structure for the flow in Fig. 9.

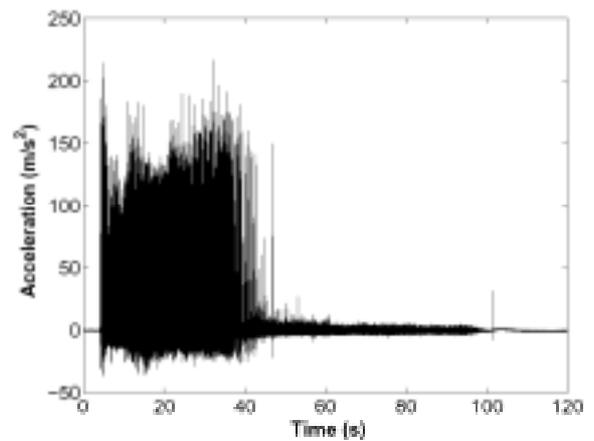

**Fig. 11a:** Vertical acceleration measurements when the accelerometer was embedded in the crushed glass. The aluminum silo had a 1.3 cm orifice and was mounted on a 31 N/mm spring. Indicated on the figure are; region I – no flow, region II – flow with pulsations and region III – flow after pulsations have ended.

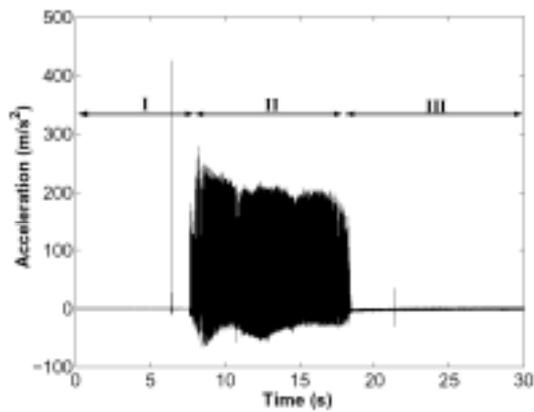

**Fig. 11b:** Vertical acceleration measurements when the accelerometer was embedded in the glass beads. The aluminum silo had a 1.9 cm orifice and was mounted on a 22 N/mm spring. Indicated on the figure are; region I – no flow, region II – flow with pulsations and region III – flow after pulsations have ended

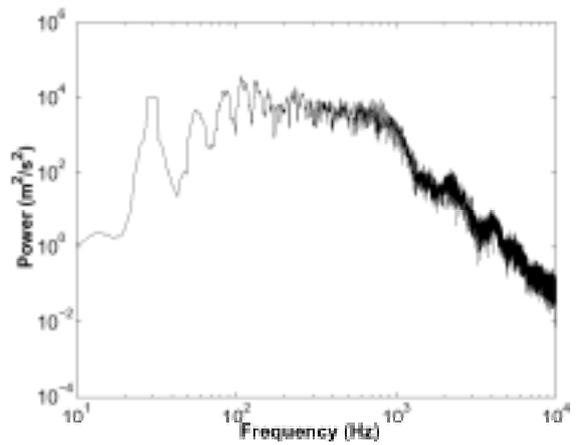

**Fig. 13a:** Power spectrum for the 20$^{th}$ second of the measurements in Fig. 11a. The power spectrum has been averaged over 4 points in frequency to make average trends clearer.

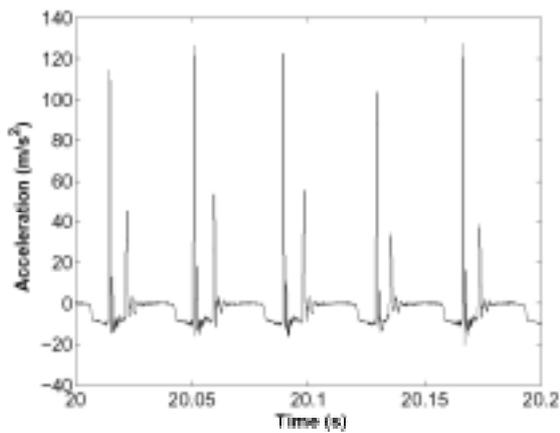

**Fig. 12a:** Close-up showing individual pulsations measured by the accelerometer in the granular material for the flow in Fig. 11a.

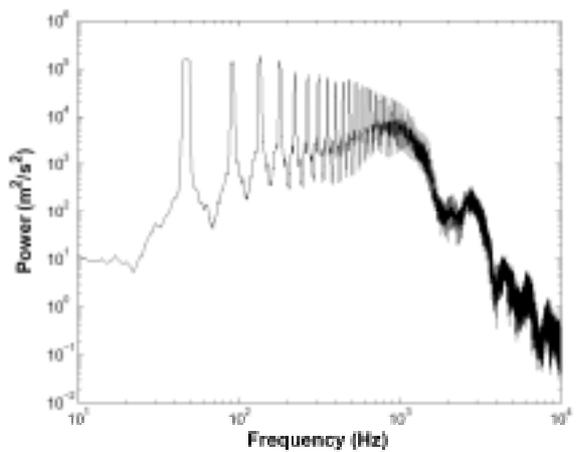

**Fig. 13b:** Power spectrum for the 15$^{th}$ second of the measurements in Fig. 11b. The power spectrum has been averaged over 4 points in frequency to make average trends clearer.

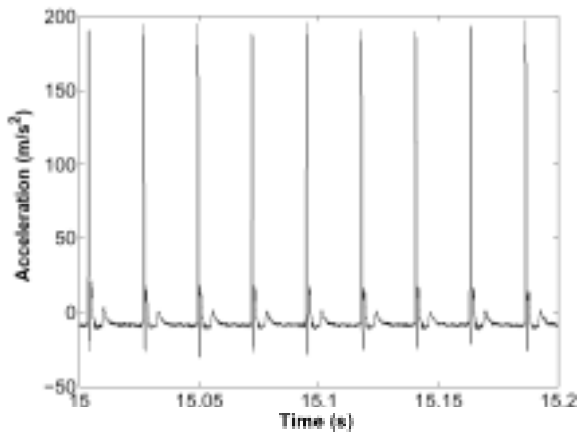

**Fig. 12b:** Close-up showing individual pulsations measured by the accelerometer in the granular material for the flow in Fig. 11b.

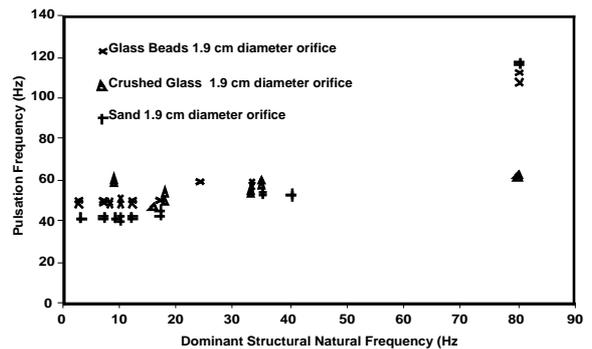

**Fig. 14:** Dominant structural natural frequency against pulsation frequency for granular materials discharging from the acrylic tube.

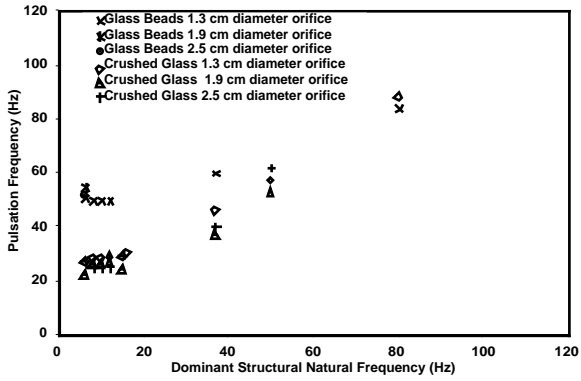

**Fig. 15:** Dominant structural natural frequency against pulsation frequency for granular materials discharging from the aluminum tube.

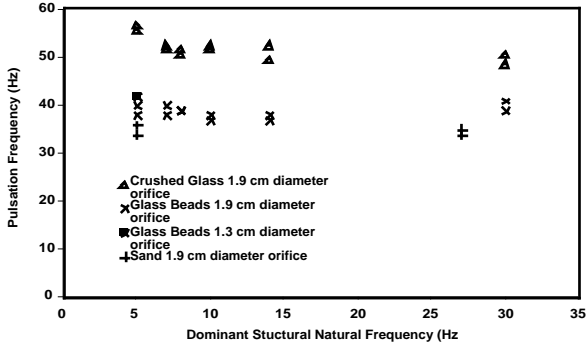

**Fig. 16:** Dominant structural natural frequency against pulsation frequency for granular materials discharging from the galvanized steel tube.

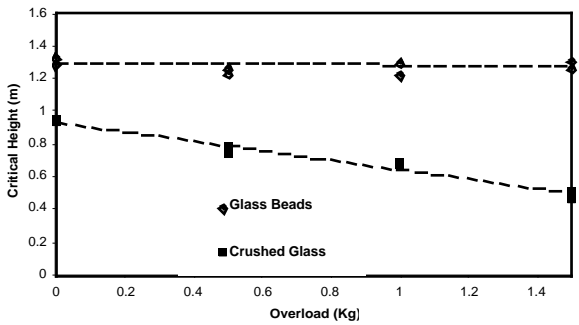

**Fig. 17:** Variation of critical height with overload for glass beads and crushed glass in the aluminum tube